\documentclass[sigconf]{acmart}

\AtBeginDocument{%
  }

\copyrightyear{2025}
\acmYear{2025}
\setcopyright{cc}
\setcctype{by}
\acmConference[CIKM '25]{Proceedings of the 34th ACM International Conference on Information and Knowledge Management}{November 10--14, 2025}{Seoul, Republic of Korea}
\acmBooktitle{Proceedings of the 34th ACM International Conference on Information and Knowledge Management (CIKM '25), November 10--14, 2025, Seoul, Republic of Korea}\acmDOI{10.1145/3746252.3760895}
\acmISBN{979-8-4007-2040-6/2025/11}

\usepackage{latexsym}
\usepackage{amsmath} 
\usepackage{colortbl}
\usepackage{booktabs}   
\usepackage{multirow}   
\usepackage{arydshln}
\usepackage{graphicx} 

\usepackage[T1]{fontenc}

\usepackage[utf8]{inputenc}

\usepackage{microtype}

\usepackage{enumitem}

\usepackage{booktabs}   
\usepackage{subcaption} 

\definecolor{lavender}{HTML}{E8E8FF}

\begin{document}

\title{Upcycling Candidate Tokens of Large Language Models \\for Query Expansion}


\author{Jinseok Kim}
\affiliation{%
  \institution{Seoul National University}
  \city{Seoul}
  \country{Republic of Korea}
}
\email{jsk0821@bdai.snu.ac.kr}

\author{Sukmin Cho}
\affiliation{%
  \institution{KAIST}
  \city{Daejeon}
  \country{Republic of Korea}}
\email{smcho@casys.kaist.ac.kr}

\author{Soyeong Jeong}
\affiliation{%
  \institution{KAIST}
  \city{Daejeon}
  \country{Republic of Korea}
}
\email{starsuzi@kaist.ac.kr}

\author{Sangyeop Kim}
\affiliation{%
 \institution{Seoul National University}
 \city{Seoul}
 \country{Republic of Korea}}
\additionalaffiliation{%
 \institution{Coxwave}
 \city{Seoul}
 \country{Republic of Korea}}

\email{sy917kim@bdai.snu.ac.kr}

\author{Sungzoon Cho}
\authornote{Corresponding author.}
\affiliation{%
  \institution{Seoul National University}
  \city{Seoul}
  \country{Republic of Korea}}
\email{zoon@snu.ac.kr}

\renewcommand{\shortauthors}{Jinseok Kim, Sukmin Cho, Soyeong Jeong, Sangyeop Kim, and Sungzoon Cho}

\begin{abstract}
Query Expansion (QE) improves retrieval performance by enriching queries with related terms. Recently, Large Language Models (LLMs) have been used for QE, but existing methods face a trade-off: generating diverse terms boosts performance but increases computational cost. To address this challenge, we propose Candidate Token Query Expansion (CTQE), which extracts diverse and relevant terms from a single LLM decoding pass by leveraging unselected candidate tokens. These tokens, though not part of the final output, are conditioned on the full query and capture useful information. By aggregating them, CTQE achieves both relevance and diversity without extra inference, reducing overhead and latency. Experiments show that CTQE delivers strong retrieval performance with significantly lower cost, outperforming or comparable to more expensive methods. Code is available at: \url{https://github.com/bluejeans8/CTQE}
\end{abstract}





\maketitle

\section{Introduction}
Information Retrieval (IR) is the task of retrieving query-relevant documents from a large corpus, playing a crucial role in diverse applications \cite{karpukhin-etal-2020-dense, product-19, trec-covid-21, legal-retrieval-22}. 
However, since queries are typically short and may lack sufficient contextual information, a key challenge in IR is bridging the lexical and semantic gap between the user’s query and the relevant documents.
To address this, Query Expansion (QE) has emerged as a promising solution by adding additional terms to the original query \cite{qe-91, prf-02}.
Recently, the powerful generation capabilities of Large Language Models (LLMs) \cite{gpt3-20, llama-23} have significantly advanced QE, substantially enhancing retrieval performance \cite{jagerman2023queryexpansionpromptinglarge, cross_qe-24, hyde-23, qe-fail-24}.

\begin{figure}
    \centering
    \includegraphics[width=0.9\columnwidth]{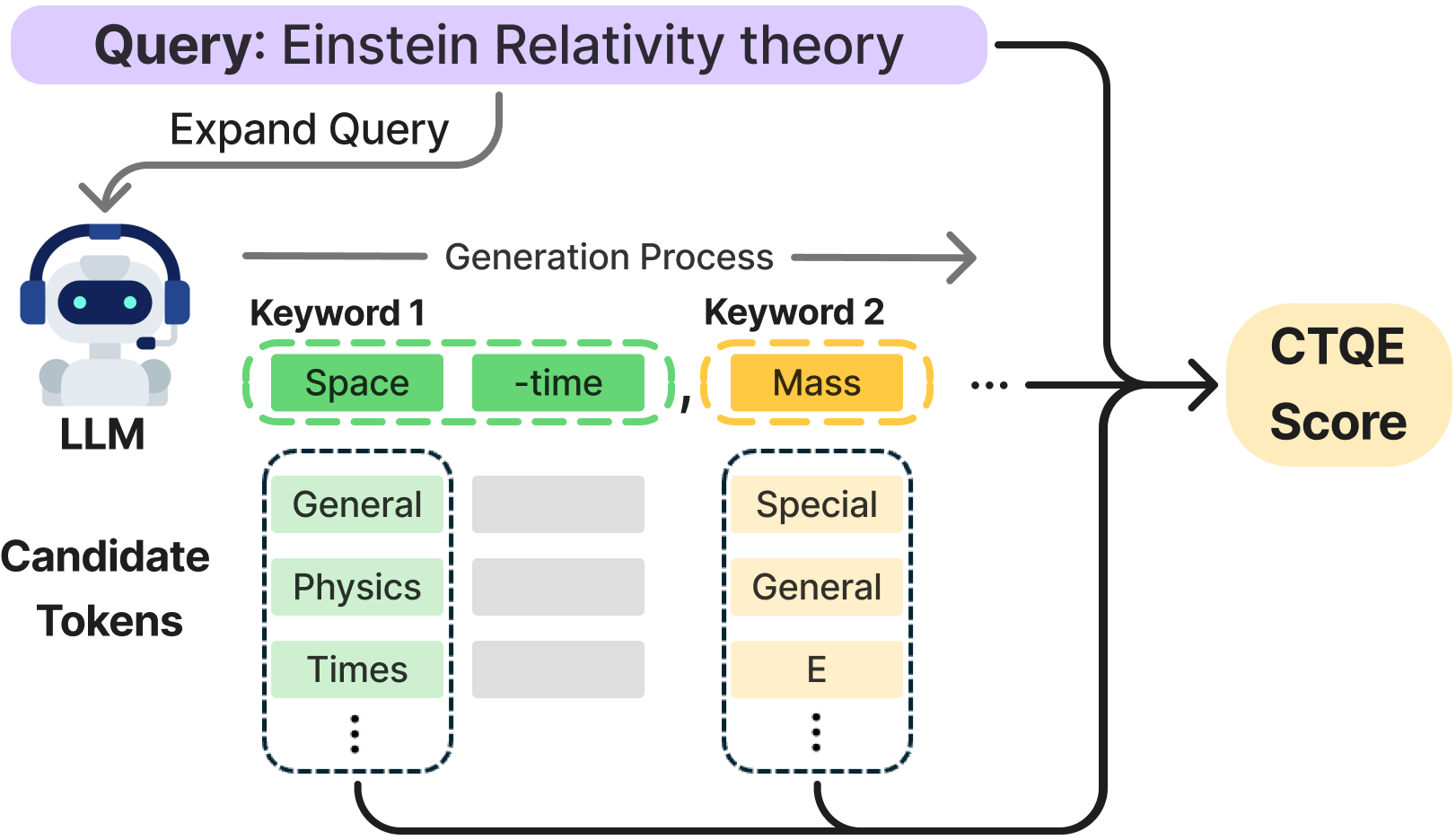}
\caption{\small An illustration of our CTQE method.} \label{fig:method}
    \vspace{-0.25in}
\end{figure}

Note that, within this QE paradigm, a key factor for improving effectiveness is the diversity of expansion keywords, as more diverse keywords can capture a broader range of semantic aspects of the user’s intent, increasing the likelihood of retrieving relevant documents that use different vocabulary or phrasing.
For example, an LLM may be prompted with the instruction: ``Write keywords that are closely related to the given query''.
In response, the LLM typically generates several keywords, each capturing semantic facets of the query, which are then appended to the original query.

To enhance the diversity of the expanded terms, researchers have explored various strategies, including using multiple manually crafted prompt variations~\cite{GenQRensemble-24, GRF-23}, prompting the model to generate long pseudo-documents~\cite{query2doc-23, jagerman2023queryexpansionpromptinglarge}, and sampling multiple generations from a single prompt using temperature scaling~\cite{mugi-24, lamer-24}.
However, while these techniques can improve diversity, they also significantly increase the computational cost, making them less practical.

To mitigate the computational overhead of previous expansion approaches, we focus on enhancing keyword-level diversity within a single decoding pass. We propose \textbf{Candidate Token Query Expansion (CTQE)}, motivated by the fact that LLMs compute probability distributions over the entire vocabulary at each step while only selecting the highest-probability token for output. Since unselected candidates are also conditioned on the query context, they can serve as valuable expansion signals. CTQE leverages these already-computed tokens to generate diverse, relevant keywords without additional inference cost. Figure~\ref{fig:method} illustrates the overall retrieval framework with CTQE, where generated keywords and candidate tokens are used alongside the original query.

We validate CTQE on 10 benchmarks using a range of LLMs with varying sizes, and evaluate performance on both sparse and dense retrievers. 
Experimental results reveal that CTQE significantly outperforms existing keyword-based query expansion methods and achieves competitive (or even outperforming) performance compared to both multi-sample keyword generation and document-level expansion, while consuming substantially fewer tokens.

\section{Methodology}
\begin{table*}[t]
\centering
\footnotesize
\setlength{\tabcolsep}{2pt} 

\renewcommand{\arraystretch}{0.95}
\begin{tabular}{lccccccccccccccccc}
\toprule
 & &\multicolumn{4}{c}{\textbf{High Resource}} 
  & \multicolumn{11}{c}{\textbf{Low Resource}}\\
\cmidrule(lr){3-6}\cmidrule(lr){7-17}

\textbf{Method} &  \textbf{PRF} & \textbf{DL19} & \textbf{DL20} & \textbf{Tokens} & \textbf{Latency} & \textbf{Covid} & \textbf{NFCorpus} & \textbf{Scifact} & \textbf{DBPedia} & \textbf{FiQA} & \textbf{Arguana} & \textbf{News} & \textbf{Robust04} & \textbf{BEIR Avg.} & \textbf{Tokens} & \textbf{Latency}\\
\midrule
\multicolumn{11}{l}{\textit{Lexical Retrieval}}\\
        BM25 & -- & 50.6	& 48.0 & -- & 0.2 &59.5 & 32.2 & 67.9 & 31.8 & 23.6 & 39.7 & 39.5 & 40.7 & 41.9 & -- & 0.1 \\
        
        \hspace{1em} + Q2D & -- & 67.0	& 64.7  & 118.2 & 3.6 & 71.1 &	36.2&	72.6&	39.7&	26.3&	39.9&	46.7&	50.2&	47.8  & 126.1 & 2.8  \\

        \hspace{1em} + MUGI & -- & 68.4 &	65.6& 591.4	& 3.9 &69.0&	36.8&	73.5&	41.0&	25.6&	35.3&	46.7&	50.1& 47.2 & 629.4 & 3.4 \\
        
    \rowcolor[HTML]{E8E8FF} 
        \hspace{1em} + Q2K & -- &58.2	&54.7 & 16.0 & 1.2 &71.6	&37.6&	70.3&	36.9&	26.5&	40.3&	47.9&	51.5&	47.8& 32.0 & 1.3 \\
    \rowcolor[HTML]{E8E8FF} 
        \hspace{1em} \textbf{+ CTQE (Ours)} & -- & \textbf{61.2}	&\textbf{56.0}& \textbf{16.0} & \textbf{1.6} &\textbf{73.8}	&\textbf{38.2}&	\textbf{70.9}&	\textbf{37.4}&	\textbf{26.9}	&\textbf{40.8}&	\textbf{49.2}&	\textbf{53.5}&	\textbf{48.8}& \textbf{32.0} & \textbf{1.4} \\

        \noalign{\vskip 0.25ex}\cdashline{1-17}\noalign{\vskip 0.75ex}

        \hspace{1em} + RM3 & $\checkmark$ & 52.2	&49.0 & -- & 0.2 & 59.3 & 34.6 & 64.6 & 30.8 & 19.2 & 38.0 & 42.6 & 42.6 & 41.4 & -- & 0.1 \\

        \hspace{1em} + Q2D/PRF & $\checkmark$ & 64.7 &	65.3 & 117.4 & 3.7 & 	75.3&	37.9&	72.7&	38.5&	29.4& 40.2	&	50.2&	51.8& 49.5 &124.5 & 3.2 \\ 

        \hspace{1em} + LameR & $\checkmark$ & 66.2	& 66.7	& 581.5 & 4.1 & 74.3 &	39.4&	73.9&	38.9&	29.4&	40.8&	49.9&	53.1& 49.9 & 623.1 & 3.4  \\
        
                
        \rowcolor[HTML]{E8E8FF} 
        \hspace{1em} + Q2K/PRF & $\checkmark$ & 61.0 &	58.1& 16.0 & 1.3 & 72.4&	38.1&	72.2&	36.8&	27.2& 40.6 &		48.8&	51.6 & 48.4 &31.8 & 1.6 \\ 
        \rowcolor[HTML]{E8E8FF} 
        \hspace{1em} \textbf{+ CTQE (Ours)} &$ \checkmark$ & \textbf{65.0}	& \textbf{62.8} & \textbf{16.0} & \textbf{1.8} & \textbf{77.8} &	\textbf{38.8}&	\textbf{73.4}&	\textbf{38.0}&	\textbf{28.0}&	\textbf{41.1}&	\textbf{50.7}&	\textbf{53.4}&	\textbf{50.2} &\textbf{31.8} & \textbf{1.7} \\


\midrule
\multicolumn{11}{l}{\textit{Neural Retrieval}}\\
        DPR$^\dagger$ & -- & 62.2	& 65.3 & -- & -- & 33.2 & 18.9 & 31.8 & 26.3 & 29.5 & 17.5 & 16.1 & 25.2 & 24.8 & -- & -- \\
        ANCE$^\dagger$ & -- & 64.5	& 64.6 & -- & -- & 65.4 & 23.7 & 50.7 & 28.1 & 30.0 & 41.5 & 38.2 & 39.2 & 39.6 & -- & -- \\
        Contriever-FT$^\dagger$ & -- & 62.1 & 63.2 & -- & -- & 59.6 & 32.8 & 67.7 & 41.3 & 32.9 & 44.6 & 42.8 & 47.3 & 46.1 & -- & -- \\

        \noalign{\vskip 0.25ex}\cdashline{1-17}\noalign{\vskip 0.75ex}

        SPLADE++ & -- & 73.1 & 72.0 & -- & 0.3 & 72.7 & 34.7 & 70.4 & 43.7 & 34.7 & 52.0 & 41.5 & 46.8 & 49.6 & -- & 0.1 \\
        \rowcolor[HTML]{E8E8FF} 
        \hspace{1em} + Q2K/PRF & $\checkmark$ & 75.5 &	74.8 & 16.0 & 1.5 & 77.9	&36.9 &	72.9 &	46.4 &	37.0 &	53.3 &	46.5 &	51.3 &	52.8& 31.8 & 1.6 \\
        \rowcolor[HTML]{E8E8FF} 
        \hspace{1em} \textbf{+ CTQE (Ours)} & $\checkmark$ & \textbf{76.9} &	\textbf{76.2}& \textbf{16.0} & \textbf{1.8} & \textbf{79.3}	&\textbf{38.0}&	\textbf{74.6}&	\textbf{46.5}&	\textbf{37.4}&	\textbf{53.5}&	\textbf{47.5}&	\textbf{54.1}&	\textbf{53.9}& \textbf{31.8} & \textbf{1.7} \\

        \noalign{\vskip 0.25ex}\cdashline{1-17}\noalign{\vskip 0.75ex}

        BGE-base& -- & 70.2 &	67.7 & -- & 2.1 & 78.2 &	37.4&	74.1&	40.7&	40.7&	45.6&	44.2&	44.4&	50.6 & -- & 0.1 \\
        \rowcolor[HTML]{E8E8FF} 
        \hspace{1em} + Q2K/PRF & $\checkmark$ &72.8	&70.3 & 16.0 & 3.4 & 82.7 &	39.2 &	76.7 &	44.0 &	42.0 &	46.0 &	47.1 &	48.1 &	53.2& 31.8 & 1.6\\
        \rowcolor[HTML]{E8E8FF} 
        \hspace{1em} \textbf{+ CTQE (Ours)} & $\checkmark$ & \textbf{73.6} &	\textbf{71.8}& \textbf{16.0} & \textbf{3.5}  & \textbf{86.1} &	\textbf{39.8} &	\textbf{77.4} &	\textbf{45.0} &	\textbf{42.9} &	\textbf{46.0} &	\textbf{48.3} &	\textbf{50.1} &	\textbf{54.4}& \textbf{31.8} & \textbf{1.6}\\

\bottomrule
\end{tabular}
\caption{\small NDCG@10 on TREC and 8 low resource datasets from BEIR. Keyword generation methods are represented by  \colorbox{lavender}{colored cells}, while our CTQE method, which additionally leverages candidate tokens, is emphasized in \textbf{bold}. $^\dagger$ represents cited results. We also report the average number of LLM output tokens and end-to-end retrieval latency (s) for both high-resource and low-resource datasets.}
    \vspace{-0.15in}
\label{tab:beir-results}
\end{table*}

\subsection{Preliminaries}
\noindent\textbf{Information Retrieval. }
We begin by defining the standard retrieval setup, in which a query \(\boldsymbol{q}\) is used to rank documents \(d \in \mathcal{D}\) based on a relevance score \(S(\boldsymbol{q}, d)\), with the goal of placing more relevant documents higher in the ranking.

\noindent\textbf{Keyword-level Query Expansion. }
To improve effectiveness, we can expand the original query \(\boldsymbol{q}\) with the LLM-generated keywords, $\boldsymbol{W} = \{\boldsymbol{w}_1, \boldsymbol{w}_2, \ldots, \boldsymbol{w}_m\}$. 
Here, each keyword \(\boldsymbol{w}_i\) is a sequence of tokens \(w_{i,1} \| \dots \| w_{i,j}\), where \(\|\) denotes sequence concatenation.
However, while increasing the diversity of generated keywords can improve retrieval performance, it requires longer output lengths or multiple decoding passes, increasing computational costs.

\subsection{Candidate Token Query Expansion (CTQE)}

To address this, we propose \textbf{CTQE}, which exploits candidate tokens considered by LLMs during decoding but not selected in the final output. At each position \(j\) of a keyword \(\boldsymbol{w}_i\), the LLM evaluates multiple alternatives before choosing the final token \(w_{i,j}\). We denote the set of the top-\(k\) alternatives as:
\[
\mathcal{C}'_{i,j} = \{w'_{i,j,1}, w'_{i,j,2}, \dots, w'_{i,j,k}\},
\]
where \(k\) is a hyperparameter specifying how many high-probability candidate tokens are retained.
To effectively utilize \(\mathcal{C}'_{i,j}\), we first filter the candidate tokens and then integrate them with the original query and generated keywords, as described in the following paragraphs.

\noindent\textbf{Token filtering. }
Since candidate tokens in \(\mathcal{C}'_{i,j}\) for \(j > 1\) are strongly conditioned on the first token \(w_{i,1}\), they tend to offer limited semantic diversity. 
To encourage diversity among candidate tokens, we collect all tokens generated at the first position of each keyword (i.e., \(\mathcal{C}'_{i,1}\)), remove duplicates, and discard tokens shorter than two characters, where the resulting set of filtered candidate tokens is:
\[
\boldsymbol{C} = \bigcup_{i=1}^{m} \left\{ w \in \mathcal{C}'_{i,1} \;\middle|\; |w| \ge 2 \right\}
\]
We next describe how these filtered candidate tokens are incorporated into query expansion.

\noindent\textbf{CTQE on Lexical Retrievers. }
For the lexical retrievers, we first construct a keyword-augmented query by appending the generated keywords to the original query: $\boldsymbol{q}_{\text{expan}} = \text{concat}(\boldsymbol{q}, \boldsymbol{W})$\footnote{
Previous studies have shown that the original query often has a greater impact on retrieval effectiveness than the generated expansion terms \cite{query2doc-23, jagerman2023queryexpansionpromptinglarge}. To reinforce the importance of the original query, we replicate $\boldsymbol{q}$ five times before concatenation. That is, we compute $\boldsymbol{q}_{\text{expan}} = \text{concat}(\boldsymbol{q}, \dots, \boldsymbol{q}, \boldsymbol{W})$ where $\boldsymbol{q}$ appears 5 times.}.
The query \(\boldsymbol{q}_{\text{expan}}\) is then used to retrieve documents by computing a relevance score \(S_{\text{expan}}(d)\) with a standard lexical retriever such as BM25 \cite{bm25}.

To further incorporate the candidate tokens \(\boldsymbol{C}\), we construct a separate subword index, and compute a candidate-token score \(S_{C}(d)\) by treating \(\boldsymbol{C}\) as a query over this index.
The final CTQE score is computed by interpolating the two signals\footnote{To ensure score compatibility, we divide $S_{\text{expan}}(d)$ by the query repetition factor (5) before interpolation.}:
\[
S_{\text{CTQE}}(d) = \alpha \cdot S_{\text{expan}}(d) + (1 - \alpha) \cdot S_{C}(d),
\]
where \(\alpha \in [0, 1]\) is a hyperparameter balancing the influence of keyword-based and candidate-based components.

\noindent\textbf{CTQE on Neural Retrievers.} 
For neural retrievers, we apply CTQE by constructing a unified query representation that combines embedding vectors from the original query, generated keywords, and candidate tokens, which is then used for document scoring, following \citet{mackie2023generativepseudorelevantfeedbacksparse}.

\noindent\textit{\textbf{Dense retrieval.}}  
For dense retrievers, we first construct a unified query representation \(\hat{v}_{\text{CTQE}}\) by linearly combining the dense representations of each query component:
\[
\hat{v}_{\text{CTQE}} = \alpha_q \cdot \hat{v}_q + \alpha_W \cdot \hat{v}_W + \alpha_C \cdot \hat{v}_C,
\]
where \(\hat{v}_q\), \(\hat{v}_W\), and \(\hat{v}_C\) denote the dense representations of the original query, generated keywords, and candidate tokens, respectively, each encoded via the query encoder, and \(\alpha_q\), \(\alpha_W\), and \(\alpha_C\) are their weights.  
The final CTQE score for \(d\) is then computed as the inner product between the fused query vector and the document vector:
\[
S_{\text{CTQE}}(d) = \hat{v}_{\text{CTQE}}^\top \hat{v}_d.
\]

\noindent\textit{\textbf{Learned sparse retrieval.}}
For learned sparse retrievers such as SPLADE \cite{splade-21}, we extend the CTQE framework by linearly combining token-level importance scores derived from each component. 
Specifically, let $\mathrm{LS}(t|\cdot)$ denote the importance score of token $t$ computed from each source using the learned sparse retriever. 
Then the final importance score for token $t$ is computed as:
\[
\mathrm{LS}_{\text{CTQE}}(t) = 
\beta_q \cdot \mathrm{LS}(t|\boldsymbol{q}) + 
\beta_W \cdot \mathrm{LS}(t|\boldsymbol{W}) + 
\beta_C \cdot \mathrm{LS}(t|\boldsymbol{C}),
\]
where $\beta_q$, $\beta_W$, and $\beta_C$ are the weights for each component\footnote {To avoid overemphasizing dominant tokens, we set \(\beta_C = 0\) for the top-20 tokens that already received the highest scores from the query and keyword components.}.
Finally, the CTQE score for $d$ is computed by matching the fused query weights against the document representation:
\[
S_{\text{CTQE}}(d) = \sum_{t \in \mathcal{V}} \mathrm{LS}_{\text{CTQE}}(t) \cdot \mathrm{LS}(t \mid d),
\]
where $\mathcal{V}$ is the total vocabulary space and $\mathrm{LS}(t \mid d)$ denotes the term weight of token $t$ in document $d$.

\section{Experimental Setups}
\noindent \textbf{Datasets and Metric. }
We evaluate CTQE on two high-resource web search datasets with large-scale annotations from TREC DL~\cite{trec19, trec20} and eight low-resource datasets from the BEIR benchmark~\cite{thakur2021beir}. As the evaluation metric, we report nDCG@10 scores.

\noindent \textbf{Baselines. }
For lexical retrieval, we include \textbf{BM25} as the base retriever.
As a classical query expansion baseline, we evaluate \textbf{RM3}~\cite{rf-01}, which leverages pseudo-relevance feedback (PRF) to refine the original query using top-ranked documents.
For LLM-based query expansion, we consider \textbf{Q2D}~\cite{jagerman2023queryexpansionpromptinglarge}, which generates hypothetical documents; \textbf{MUGI}~\cite{mugi-24}, which produces multiple pseudo-references and reweights them adaptively; \textbf{Q2K}, which generates keyword-based expansions; and \textbf{LameR} \cite{lamer-24}, which generates multiple pseudo-documents conditioned on the retrieved documents.
Additionally, we consider PRF variants \textbf{Q2D/PRF} and \textbf{Q2K/PRF}, where the LLM incorporates initially retrieved results for refinement. 

For neural retriever, we include a recent dense model, \textbf{BGE-base}\footnote{ \texttt{BAAI/bge-base-en-v1.5}}\cite{bge_embedding} and learned sparse retriever \textbf{SPLADE++}\footnote{\texttt{naver/splade-cocondenser-ensembledistil} }~\cite{splade-pp-22}.
We also evaluate \textbf{Q2K/PRF} variants, where keyword-based query expansion is applied to these neural retrievers.
Additionally, we report results from three widely used dense retrievers for further reference: \textbf{DPR}~\cite{karpukhin-etal-2020-dense}, \textbf{ANCE}~\cite{ance-20}, and \textbf{Contriever-FT}~\cite{contriever-21}.

\begin{figure*}[t!]
    \begin{minipage}{0.29\linewidth}
        \centering
        \includegraphics[width=0.95\columnwidth]{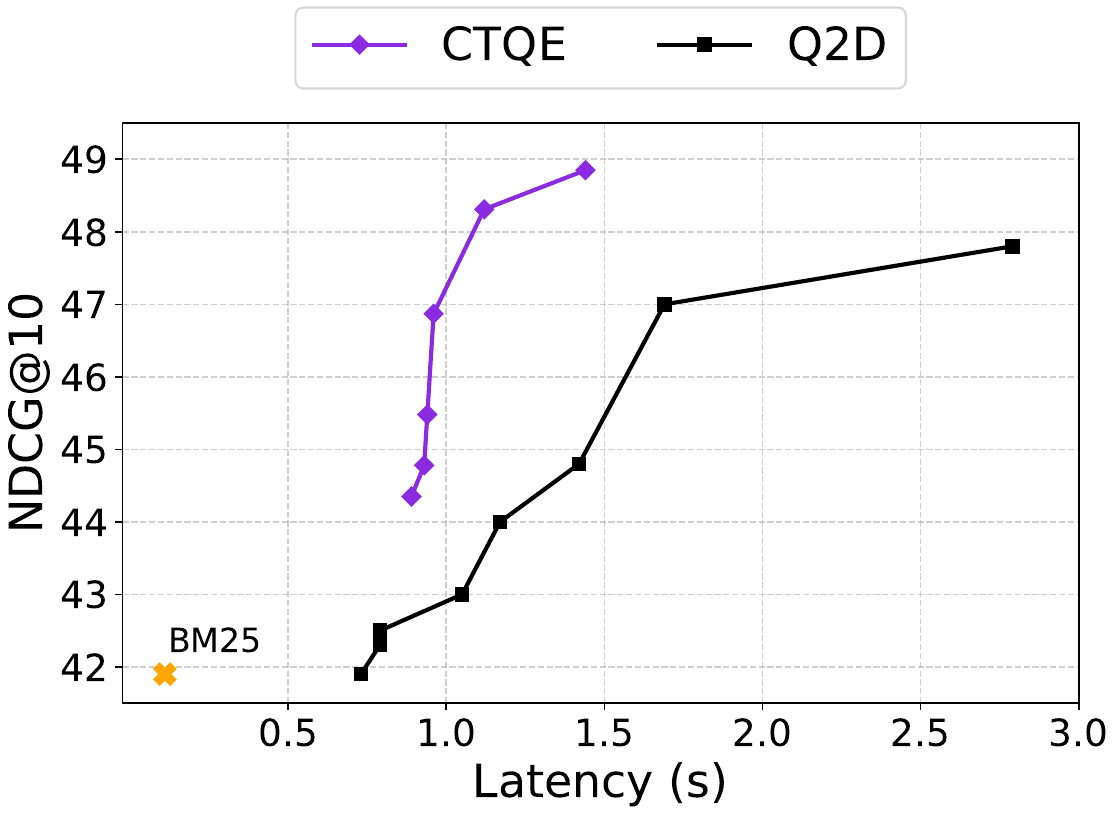}
    \end{minipage}
    \begin{minipage}{0.4\linewidth}
        \centering
        \includegraphics[width=0.95\columnwidth]{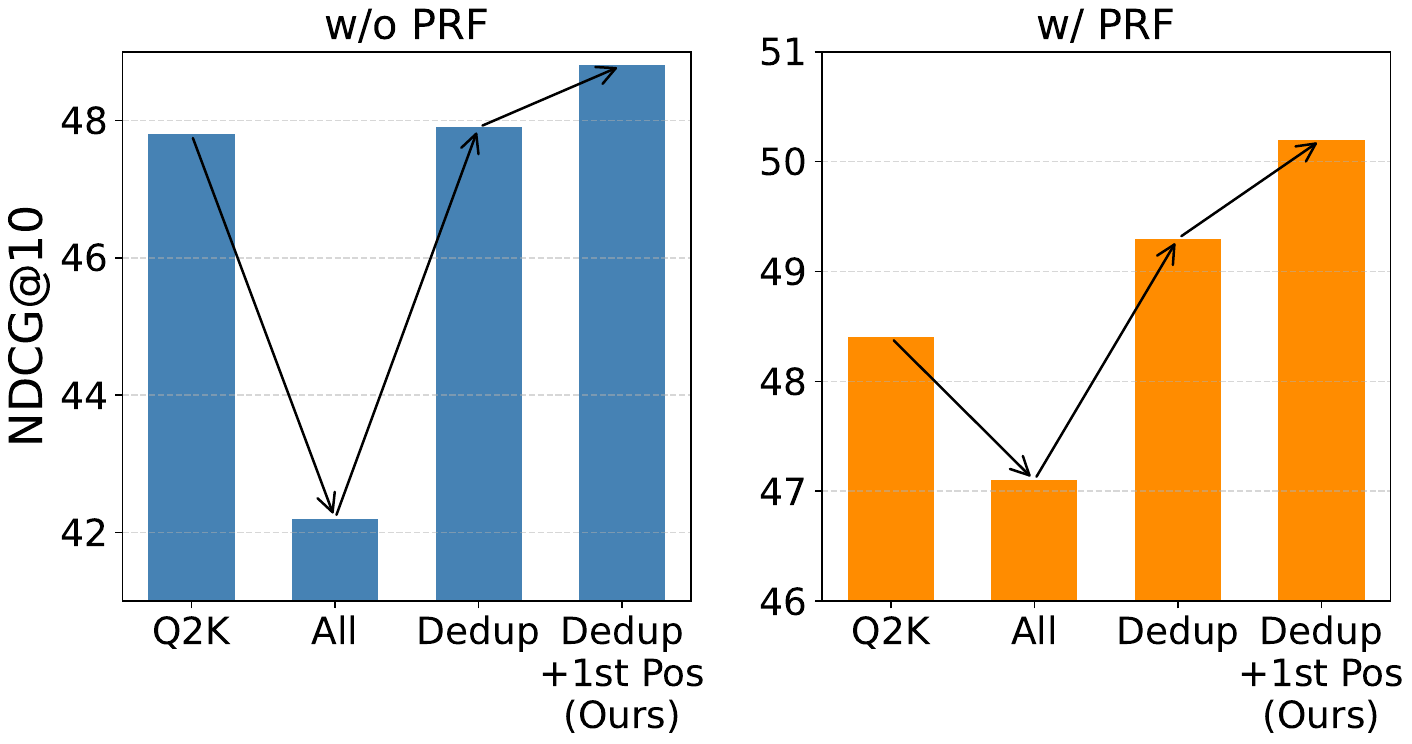}
    \end{minipage}        
    \begin{minipage}{0.29\linewidth}
        \centering
        \includegraphics[width=0.95\columnwidth]{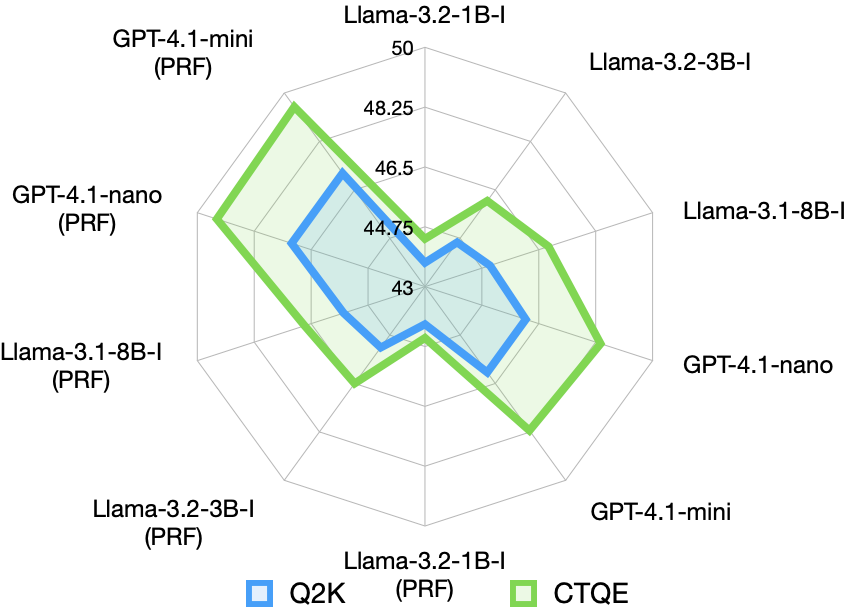}
    \end{minipage}
    
    \vspace{-0.125in}
    \caption{\small (Left). Retrieval performance vs. latency for CTQE (w/o PRF) and Q2D, averaged across 8 low-resource datasets. Each point corresponds to a different maximum generation length. CTQE is evaluated at token limits of [1, 2, 4, 8, 16, 32], while Q2D uses [1, 2, 4, 8, 16, 32, 64, 128]. (Center) Average NDCG@10 scores for incremental candidate token filtering on eight low-resource datasets. (Right). Retrieval performance for different LLMs (Llama-3 and GPT-4.1 family) averaged over 8 low-resource datasets, using a maximum generation length of 16.}
    \label{fig:classifier}
    \vspace{-0.15in}
\end{figure*}

\noindent \textbf{Implementation Details. } 
For the base LLMs, we incorporate diverse GPT-4.1~\cite{openai2024gpt41} series and the LLaMA 3~\cite{Llama3} series models, but mainly report results using \texttt{gpt-4.1-mini}.
Specifically, we set the max token length to 16 for high-resource datasets and 32 for low-resource ones when generating keywords with temperature 0. We set \(k=20\), corresponding to the top-20 candidate tokens considered at each decoding step, the maximum provided by the OpenAI API.

For a fair comparison, we adopt a uniform, task-agnostic zero-shot prompt~\cite{jagerman2023queryexpansionpromptinglarge} across all pseudo-document generation baselines, generating up to 128 tokens following~\cite{query2doc-23}. 
For Q2K, we prompt the model with the instruction: \textit{"Write keywords that are closely related to the given query."} 
For the Q2K/PRF variant, we additionally provide the top retrieved passages as contextual input. 
PRF-based methods use the top-10 documents retrieved by BM25, each truncated to a maximum of 128 tokens~\cite{lamer-24}, and multi-text generation methods sample 5 generations with a temperature of 1.0.

Regarding the hyperparameter, we set $\alpha=0.9$ for score interpolation for the lexical retriever. 
For neural retrievers, we set $\alpha_q=0.5$, $\alpha_W=0.1$, $\alpha_C=0.1$ for dense retrieval, and $\beta_q=0.5$, $\beta_W=0.1$, $\beta_C=0.1$ for learned sparse retrieval. All experiments are implemented using the Pyserini toolkit \cite{pyserini-21} with a cloud VM equipped with a single NVIDIA A100 GPU (40GB).

\section{Experimental Results and Analyses}
\noindent \textbf{Overall Results.}
As shown in Table \ref{tab:beir-results}, CTQE consistently outperforms the keyword-based expansion strategy (Q2K) across all datasets, demonstrating the effectiveness of upcycling candidate tokens for query expansion.
Specifically, our method achieves comparable results to pseudo-document generation approaches, such as Q2D and LameR, with a significantly lower token budget, resulting in much faster execution.
Surprisingly, CTQE is particularly effective on low-resource datasets, even outperforming the pseudo-document generation approaches, suggesting that candidate tokens often capture domain-relevant terms that enhance retrieval performance.
Furthermore, CTQE with BM25 outperforms several strong dense retrievers, including DPR, ANCE, and Contriever-FT, which require substantial fine-tuning, thereby underscoring its strong zero-shot capabilities. Lastly, CTQE is not only effective for sparse retrievers, but also integrates seamlessly with powerful neural models such as SPLADE and BGE, demonstrating its robustness and generalizability.

\noindent\textbf{Performance vs Latency. }
We further explore the trade-off between retrieval performance and latency for Q2D and CTQE across varying token lengths in the left plot. 
As shown in the left plot in Figure \ref{fig:classifier}, we observe that Q2D performs poorly with shorter outputs, indicating that generating pseudo-documents requires a sufficient number of tokens to contain informative content for effective retrieval. However, this naturally leads to increased latency as more tokens are generated.

In contrast, CTQE achieves strong performance with minimal token usage, demonstrating significantly lower latency. Remarkably, even with a single token, CTQE outperforms BM25 by more than 2 points in NDCG@10. This highlights the efficiency and effectiveness of CTQE in resource-constrained environments, as it can provide high retrieval performance with minimal computational cost.

\noindent\textbf{Comparing Keyword Methods. }
Table \ref{tab:keyword-gen} highlights the trade-offs between keyword diversity, token usage, and retrieval performance across different keyword generation strategies.
In CTQE, keyword diversity is measured by counting both the generated keywords and the filtered candidate tokens set.

Methods using multiple prompt variations (ensemble prompting), unconstrained generation, or explicitly prompting keyword counts ($k=10, 20, 30$) can increase keyword diversity but suffer from lower NDCG@10 scores compared to the fixed-length setting. This suggests that longer outputs tend to include less relevant or redundant keywords, negatively impacting retrieval performance. While sampling-based approaches with temperature scaling ($T=1, 2$) improve keyword diversity and outperform Q2K under fixed-length settings, they still incur the overhead of multiple decoding passes.

Our CTQE method achieves the highest diversity (33.3 unique keywords) and best performance (65.0 NDCG@10) with only a single decoding pass. By leveraging candidate tokens from each generation step, CTQE efficiently extracts additional relevant keywords without increasing token consumption, demonstrating superior efficiency in the diversity-performance trade-off.

\noindent\textbf{Effect of Token filtering. }
The center plot in Figure \ref{fig:classifier} presents an ablation study of our CTQE approach, showing filtering effects on retrieval performance. First, CTQE using all candidate tokens without any filtering (All) shows a performance drop compared to Q2K. This suggests that indiscriminately adding candidate tokens can introduce noise. When we apply duplicate removal to eliminate redundant tokens (Dedup), we observe a consistent performance improvement. Finally, limiting the expansion to candidate tokens from only the first position of each keyword ($C'_{i,1}$) (Dedup + 1st pos) yields substantial gains. These results demonstrate that first-position candidate tokens provide sufficient lexical diversity while systematic filtering reduces noise and maintains query clarity.











\begin{table}[t]
\scriptsize
\centering
\renewcommand{\arraystretch}{0.95}
\begin{tabular}{lccc}
\toprule
\textbf{Method} & \textbf{Tokens} & \textbf{Unique Keywords} & \textbf{NDCG@10} \\
\midrule

\multicolumn{4}{l}{\textit{Unconstrained Length}} \\

\hspace{1em}Ensemble-Prompting \cite{GenQRensemble-24} & 581.4 & 50.8   & 54.9  \\

\hspace{1em}Q2K (Unconstrained) & 63.4 & 17.2 & 60.9 \\

\noalign{\vskip 0.25ex}\cdashline{1-4}\noalign{\vskip 0.75ex}

\multicolumn{4}{l}{\textit{Prompting Keyword Count ($k$)}} \\
\hspace{1em}$k=10$ & 36.4 & 10 & 59.8 \\
\hspace{1em}$k=20$ & 74.9 & 20 & 58.4 \\
\hspace{1em}$k=30$ & 115.7 & 30.4 & 55.5 \\

\noalign{\vskip 0.25ex}\cdashline{1-4}\noalign{\vskip 0.75ex}

\multicolumn{4}{l}{\textit{Fixed Length (16)}} \\
\hspace{1em}Q2K & 16 & 5 & 61.0 \\

\hspace{1em}Sampling ($T=1$) & 80 & 9.1 & 61.5 \\
\hspace{1em}Sampling ($T=2$) & 80 & 12.5 & 62.9 \\

\hspace{1em}\textbf{CTQE (Ours)} & 16 & 33.3 & \textbf{65.0} \\

\bottomrule
\end{tabular}
\caption{Comparison of different keyword generation strategies under the PRF setting by token usage, keyword diversity, and retrieval performance (NDCG@10) on TREC19.}
\vspace{-0.8em}
\label{tab:keyword-gen}
\end{table}

\noindent\textbf{Impact of Different LLMs. }
To evaluate the robustness of our CTQE across diverse LLMs, we present the results of using various LLMs for keyword generation in the right plot in Figure \ref{fig:classifier}. Notably, CTQE consistently outperforms Q2K across all LLMs, demonstrating strong adaptability and flexibility.
Furthermore, CTQE effectively leverages the capabilities of stronger LLMs, such as the GPT series, as evidenced by the significantly higher gains with these models. 
These results confirm that our method is generalizable across diverse models and benefits from advancements in LLMs.

\section{Conclusion}
We presented Candidate Token Query Expansion (CTQE), a novel framework that enhances query expansion by upcycling candidate tokens discarded during LLM decoding. CTQE efficiently captures diverse and relevant lexical signals while outperforming keyword-based methods and remaining competitive with pseudo-document approaches across benchmarks. These results highlight CTQE's ability to balance retrieval effectiveness and computational cost, making it a practical and scalable solution for real-world search systems.

\section*{Acknowledgements}
This work was supported by Institute of Information \& communications Technology Planning \& Evaluation (IITP) grant funded by the Korea government(MSIT) [NO.RS-2021-II211343, Artificial Intelligence Graduate School Program (Seoul National University)]

\section*{GenAI Usage Disclosure}
The authors used ChatGPT to assist with minor grammar corrections and to help identify minor bugs in code snippets. The authors take full responsibility for the final content of the manuscript and code.

\bibliographystyle{ACM-Reference-Format}
\bibliography{references}


\begin{thebibliography}{31}


\ifx \showCODEN    \undefined \def \showCODEN     #1{\unskip}     \fi
\ifx \showISBNx    \undefined \def \showISBNx     #1{\unskip}     \fi
\ifx \showISBNxiii \undefined \def \showISBNxiii  #1{\unskip}     \fi
\ifx \showISSN     \undefined \def \showISSN      #1{\unskip}     \fi
\ifx \showLCCN     \undefined \def \showLCCN      #1{\unskip}     \fi
\ifx \shownote     \undefined \def \shownote      #1{#1}          \fi
\ifx \showarticletitle \undefined \def \showarticletitle #1{#1}   \fi
\ifx \showURL      \undefined \def \showURL       {\relax}        \fi
\providecommand\bibfield[2]{#2}
\providecommand\bibinfo[2]{#2}
\providecommand\natexlab[1]{#1}
\providecommand\showeprint[2][]{arXiv:#2}

\bibitem[Amati and Van~Rijsbergen(2002)]%
        {prf-02}
\bibfield{author}{\bibinfo{person}{Gianni Amati} {and} \bibinfo{person}{Cornelis~Joost Van~Rijsbergen}.} \bibinfo{year}{2002}\natexlab{}.
\newblock \showarticletitle{Probabilistic models of information retrieval based on measuring the divergence from randomness}.
\newblock \bibinfo{journal}{\emph{ACM Trans. Inf. Syst.}} \bibinfo{volume}{20}, \bibinfo{number}{4} (\bibinfo{date}{Oct.} \bibinfo{year}{2002}), \bibinfo{pages}{357–389}.
\newblock
\showISSN{1046-8188}
\href{https://doi.org/10.1145/582415.582416}{doi:\nolinkurl{10.1145/582415.582416}}


\bibitem[Brown et~al\mbox{.}(2020)]%
        {gpt3-20}
\bibfield{author}{\bibinfo{person}{Tom~B. Brown}, \bibinfo{person}{Benjamin Mann}, \bibinfo{person}{Nick Ryder}, \bibinfo{person}{Melanie Subbiah}, \bibinfo{person}{Jared Kaplan}, \bibinfo{person}{Prafulla Dhariwal}, \bibinfo{person}{Arvind Neelakantan}, \bibinfo{person}{Pranav Shyam}, \bibinfo{person}{Girish Sastry}, \bibinfo{person}{Amanda Askell}, \bibinfo{person}{Sandhini Agarwal}, \bibinfo{person}{Ariel Herbert-Voss}, \bibinfo{person}{Gretchen Krueger}, \bibinfo{person}{Tom Henighan}, \bibinfo{person}{Rewon Child}, \bibinfo{person}{Aditya Ramesh}, \bibinfo{person}{Daniel~M. Ziegler}, \bibinfo{person}{Jeffrey Wu}, \bibinfo{person}{Clemens Winter}, \bibinfo{person}{Christopher Hesse}, \bibinfo{person}{Mark Chen}, \bibinfo{person}{Eric Sigler}, \bibinfo{person}{Mateusz Litwin}, \bibinfo{person}{Scott Gray}, \bibinfo{person}{Benjamin Chess}, \bibinfo{person}{Jack Clark}, \bibinfo{person}{Christopher Berner}, \bibinfo{person}{Sam McCandlish}, \bibinfo{person}{Alec Radford}, \bibinfo{person}{Ilya Sutskever},
  {and} \bibinfo{person}{Dario Amodei}.} \bibinfo{year}{2020}\natexlab{}.
\newblock \showarticletitle{Language models are few-shot learners}. In \bibinfo{booktitle}{\emph{Proceedings of the 34th International Conference on Neural Information Processing Systems}} (Vancouver, BC, Canada) \emph{(\bibinfo{series}{NIPS '20})}. \bibinfo{publisher}{Curran Associates Inc.}, \bibinfo{address}{Red Hook, NY, USA}, Article \bibinfo{articleno}{159}, \bibinfo{numpages}{25}~pages.
\newblock
\showISBNx{9781713829546}


\bibitem[Craswell et~al\mbox{.}(2021)]%
        {trec20}
\bibfield{author}{\bibinfo{person}{Nick Craswell}, \bibinfo{person}{Bhaskar Mitra}, \bibinfo{person}{Emine Yilmaz}, {and} \bibinfo{person}{Daniel Campos}.} \bibinfo{year}{2021}\natexlab{}.
\newblock \bibinfo{title}{Overview of the TREC 2020 deep learning track}.
\newblock
\showeprint[arxiv]{2102.07662}~[cs.IR]
\urldef\tempurl%
\url{https://arxiv.org/abs/2102.07662}
\showURL{%
\tempurl}


\bibitem[Craswell et~al\mbox{.}(2020)]%
        {trec19}
\bibfield{author}{\bibinfo{person}{Nick Craswell}, \bibinfo{person}{Bhaskar Mitra}, \bibinfo{person}{Emine Yilmaz}, \bibinfo{person}{Daniel Campos}, {and} \bibinfo{person}{Ellen~M. Voorhees}.} \bibinfo{year}{2020}\natexlab{}.
\newblock \bibinfo{title}{Overview of the TREC 2019 deep learning track}.
\newblock
\showeprint[arxiv]{2003.07820}~[cs.IR]
\urldef\tempurl%
\url{https://arxiv.org/abs/2003.07820}
\showURL{%
\tempurl}


\bibitem[Dhole and Agichtein(2024)]%
        {GenQRensemble-24}
\bibfield{author}{\bibinfo{person}{Kaustubh~D. Dhole} {and} \bibinfo{person}{Eugene Agichtein}.} \bibinfo{year}{2024}\natexlab{}.
\newblock \showarticletitle{GenQREnsemble: Zero-Shot LLM Ensemble Prompting for Generative Query Reformulation}. In \bibinfo{booktitle}{\emph{Advances in Information Retrieval: 46th European Conference on Information Retrieval, ECIR 2024, Glasgow, UK, March 24–28, 2024, Proceedings, Part III}} (Glasgow, United Kingdom). \bibinfo{publisher}{Springer-Verlag}, \bibinfo{address}{Berlin, Heidelberg}, \bibinfo{pages}{326–335}.
\newblock
\showISBNx{978-3-031-56062-0}
\href{https://doi.org/10.1007/978-3-031-56063-7_24}{doi:\nolinkurl{10.1007/978-3-031-56063-7_24}}


\bibitem[Dubey et~al\mbox{.}(2024)]%
        {Llama3}
\bibfield{author}{\bibinfo{person}{Abhimanyu Dubey}, \bibinfo{person}{Abhinav Jauhri}, \bibinfo{person}{Abhinav Pandey}, \bibinfo{person}{Abhishek Kadian}, \bibinfo{person}{Ahmad Al{-}Dahle}, \bibinfo{person}{Aiesha Letman}, \bibinfo{person}{Akhil Mathur}, \bibinfo{person}{Alan Schelten}, \bibinfo{person}{Amy Yang}, \bibinfo{person}{Angela Fan}, \bibinfo{person}{Anirudh Goyal}, \bibinfo{person}{Anthony Hartshorn}, \bibinfo{person}{Aobo Yang}, \bibinfo{person}{Archi Mitra}, \bibinfo{person}{Archie Sravankumar}, \bibinfo{person}{Artem Korenev}, \bibinfo{person}{Arthur Hinsvark}, \bibinfo{person}{Arun Rao}, \bibinfo{person}{Aston Zhang}, \bibinfo{person}{Aur{\'{e}}lien Rodriguez}, \bibinfo{person}{Austen Gregerson}, \bibinfo{person}{Ava Spataru}, \bibinfo{person}{Baptiste Rozi{\`{e}}re}, \bibinfo{person}{Bethany Biron}, \bibinfo{person}{Binh Tang}, \bibinfo{person}{Bobbie Chern}, \bibinfo{person}{Charlotte Caucheteux}, \bibinfo{person}{Chaya Nayak}, \bibinfo{person}{Chloe Bi}, \bibinfo{person}{Chris Marra},
  \bibinfo{person}{Chris McConnell}, \bibinfo{person}{Christian Keller}, \bibinfo{person}{Christophe Touret}, \bibinfo{person}{Chunyang Wu}, \bibinfo{person}{Corinne Wong}, \bibinfo{person}{Cristian~Canton Ferrer}, \bibinfo{person}{Cyrus Nikolaidis}, \bibinfo{person}{Damien Allonsius}, \bibinfo{person}{Daniel Song}, \bibinfo{person}{Danielle Pintz}, \bibinfo{person}{Danny Livshits}, \bibinfo{person}{David Esiobu}, \bibinfo{person}{Dhruv Choudhary}, \bibinfo{person}{Dhruv Mahajan}, \bibinfo{person}{Diego Garcia{-}Olano}, \bibinfo{person}{Diego Perino}, \bibinfo{person}{Dieuwke Hupkes}, \bibinfo{person}{Egor Lakomkin}, \bibinfo{person}{Ehab AlBadawy}, \bibinfo{person}{Elina Lobanova}, \bibinfo{person}{Emily Dinan}, \bibinfo{person}{Eric~Michael Smith}, \bibinfo{person}{Filip Radenovic}, \bibinfo{person}{Frank Zhang}, \bibinfo{person}{Gabriel Synnaeve}, \bibinfo{person}{Gabrielle Lee}, \bibinfo{person}{Georgia~Lewis Anderson}, \bibinfo{person}{Graeme Nail}, \bibinfo{person}{Gr{\'{e}}goire Mialon},
  \bibinfo{person}{Guan Pang}, \bibinfo{person}{Guillem Cucurell}, \bibinfo{person}{Hailey Nguyen}, \bibinfo{person}{Hannah Korevaar}, \bibinfo{person}{Hu Xu}, \bibinfo{person}{Hugo Touvron}, \bibinfo{person}{Iliyan Zarov}, \bibinfo{person}{Imanol~Arrieta Ibarra}, \bibinfo{person}{Isabel~M. Kloumann}, \bibinfo{person}{Ishan Misra}, \bibinfo{person}{Ivan Evtimov}, \bibinfo{person}{Jade Copet}, \bibinfo{person}{Jaewon Lee}, \bibinfo{person}{Jan Geffert}, \bibinfo{person}{Jana Vranes}, \bibinfo{person}{Jason Park}, \bibinfo{person}{Jay Mahadeokar}, \bibinfo{person}{Jeet Shah}, \bibinfo{person}{Jelmer van~der Linde}, \bibinfo{person}{Jennifer Billock}, \bibinfo{person}{Jenny Hong}, \bibinfo{person}{Jenya Lee}, \bibinfo{person}{Jeremy Fu}, \bibinfo{person}{Jianfeng Chi}, \bibinfo{person}{Jianyu Huang}, \bibinfo{person}{Jiawen Liu}, \bibinfo{person}{Jie Wang}, \bibinfo{person}{Jiecao Yu}, \bibinfo{person}{Joanna Bitton}, \bibinfo{person}{Joe Spisak}, \bibinfo{person}{Jongsoo Park}, \bibinfo{person}{Joseph Rocca},
  \bibinfo{person}{Joshua Johnstun}, \bibinfo{person}{Joshua Saxe}, \bibinfo{person}{Junteng Jia}, \bibinfo{person}{Kalyan~Vasuden Alwala}, \bibinfo{person}{Kartikeya Upasani}, \bibinfo{person}{Kate Plawiak}, \bibinfo{person}{Ke Li}, \bibinfo{person}{Kenneth Heafield}, \bibinfo{person}{Kevin Stone}, {and} \bibinfo{person}{et al.}} \bibinfo{year}{2024}\natexlab{}.
\newblock \showarticletitle{The Llama 3 Herd of Models}.
\newblock \bibinfo{journal}{\emph{arXiv preprint arXiv:2407.21783}} (\bibinfo{year}{2024}).
\newblock
\href{https://doi.org/10.48550/ARXIV.2407.21783}{doi:\nolinkurl{10.48550/ARXIV.2407.21783}}
\showeprint[arXiv]{2407.21783}


\bibitem[Formal et~al\mbox{.}(2022)]%
        {splade-pp-22}
\bibfield{author}{\bibinfo{person}{Thibault Formal}, \bibinfo{person}{Carlos Lassance}, \bibinfo{person}{Benjamin Piwowarski}, {and} \bibinfo{person}{St\'{e}phane Clinchant}.} \bibinfo{year}{2022}\natexlab{}.
\newblock \showarticletitle{From Distillation to Hard Negative Sampling: Making Sparse Neural IR Models More Effective}. In \bibinfo{booktitle}{\emph{Proceedings of the 45th International ACM SIGIR Conference on Research and Development in Information Retrieval}} (Madrid, Spain) \emph{(\bibinfo{series}{SIGIR '22})}. \bibinfo{publisher}{Association for Computing Machinery}, \bibinfo{address}{New York, NY, USA}, \bibinfo{pages}{2353–2359}.
\newblock
\showISBNx{9781450387323}
\href{https://doi.org/10.1145/3477495.3531857}{doi:\nolinkurl{10.1145/3477495.3531857}}


\bibitem[Formal et~al\mbox{.}(2021)]%
        {splade-21}
\bibfield{author}{\bibinfo{person}{Thibault Formal}, \bibinfo{person}{Benjamin Piwowarski}, {and} \bibinfo{person}{St\'{e}phane Clinchant}.} \bibinfo{year}{2021}\natexlab{}.
\newblock \showarticletitle{SPLADE: Sparse Lexical and Expansion Model for First Stage Ranking}. In \bibinfo{booktitle}{\emph{Proceedings of the 44th International ACM SIGIR Conference on Research and Development in Information Retrieval}} (Virtual Event, Canada) \emph{(\bibinfo{series}{SIGIR '21})}. \bibinfo{publisher}{Association for Computing Machinery}, \bibinfo{address}{New York, NY, USA}, \bibinfo{pages}{2288–2292}.
\newblock
\showISBNx{9781450380379}
\href{https://doi.org/10.1145/3404835.3463098}{doi:\nolinkurl{10.1145/3404835.3463098}}


\bibitem[Gao et~al\mbox{.}(2023)]%
        {hyde-23}
\bibfield{author}{\bibinfo{person}{Luyu Gao}, \bibinfo{person}{Xueguang Ma}, \bibinfo{person}{Jimmy Lin}, {and} \bibinfo{person}{Jamie Callan}.} \bibinfo{year}{2023}\natexlab{}.
\newblock \showarticletitle{Precise Zero-Shot Dense Retrieval without Relevance Labels}. In \bibinfo{booktitle}{\emph{Proceedings of the 61st Annual Meeting of the Association for Computational Linguistics (Volume 1: Long Papers)}}, \bibfield{editor}{\bibinfo{person}{Anna Rogers}, \bibinfo{person}{Jordan Boyd-Graber}, {and} \bibinfo{person}{Naoaki Okazaki}} (Eds.). \bibinfo{publisher}{Association for Computational Linguistics}, \bibinfo{address}{Toronto, Canada}, \bibinfo{pages}{1762--1777}.
\newblock
\href{https://doi.org/10.18653/v1/2023.acl-long.99}{doi:\nolinkurl{10.18653/v1/2023.acl-long.99}}


\bibitem[Izacard et~al\mbox{.}(2022)]%
        {contriever-21}
\bibfield{author}{\bibinfo{person}{Gautier Izacard}, \bibinfo{person}{Mathilde Caron}, \bibinfo{person}{Lucas Hosseini}, \bibinfo{person}{Sebastian Riedel}, \bibinfo{person}{Piotr Bojanowski}, \bibinfo{person}{Armand Joulin}, {and} \bibinfo{person}{Edouard Grave}.} \bibinfo{year}{2022}\natexlab{}.
\newblock \showarticletitle{Unsupervised Dense Information Retrieval with Contrastive Learning}.
\newblock \bibinfo{journal}{\emph{Transactions on Machine Learning Research}} (\bibinfo{year}{2022}).
\newblock
\showISSN{2835-8856}
\urldef\tempurl%
\url{https://openreview.net/forum?id=jKN1pXi7b0}
\showURL{%
\tempurl}


\bibitem[Jagerman et~al\mbox{.}(2023)]%
        {jagerman2023queryexpansionpromptinglarge}
\bibfield{author}{\bibinfo{person}{Rolf Jagerman}, \bibinfo{person}{Honglei Zhuang}, \bibinfo{person}{Zhen Qin}, \bibinfo{person}{Xuanhui Wang}, {and} \bibinfo{person}{Michael Bendersky}.} \bibinfo{year}{2023}\natexlab{}.
\newblock \bibinfo{title}{Query Expansion by Prompting Large Language Models}.
\newblock
\showeprint[arxiv]{2305.03653}~[cs.IR]
\urldef\tempurl%
\url{https://arxiv.org/abs/2305.03653}
\showURL{%
\tempurl}


\bibitem[Karpukhin et~al\mbox{.}(2020)]%
        {karpukhin-etal-2020-dense}
\bibfield{author}{\bibinfo{person}{Vladimir Karpukhin}, \bibinfo{person}{Barlas Oguz}, \bibinfo{person}{Sewon Min}, \bibinfo{person}{Patrick Lewis}, \bibinfo{person}{Ledell Wu}, \bibinfo{person}{Sergey Edunov}, \bibinfo{person}{Danqi Chen}, {and} \bibinfo{person}{Wen-tau Yih}.} \bibinfo{year}{2020}\natexlab{}.
\newblock \showarticletitle{Dense Passage Retrieval for Open-Domain Question Answering}. In \bibinfo{booktitle}{\emph{Proceedings of the 2020 Conference on Empirical Methods in Natural Language Processing (EMNLP)}}, \bibfield{editor}{\bibinfo{person}{Bonnie Webber}, \bibinfo{person}{Trevor Cohn}, \bibinfo{person}{Yulan He}, {and} \bibinfo{person}{Yang Liu}} (Eds.). \bibinfo{publisher}{Association for Computational Linguistics}, \bibinfo{address}{Online}, \bibinfo{pages}{6769--6781}.
\newblock
\href{https://doi.org/10.18653/v1/2020.emnlp-main.550}{doi:\nolinkurl{10.18653/v1/2020.emnlp-main.550}}


\bibitem[Lavrenko and Croft(2001)]%
        {rf-01}
\bibfield{author}{\bibinfo{person}{Victor Lavrenko} {and} \bibinfo{person}{W.~Bruce Croft}.} \bibinfo{year}{2001}\natexlab{}.
\newblock \showarticletitle{Relevance based language models}. In \bibinfo{booktitle}{\emph{Proceedings of the 24th Annual International ACM SIGIR Conference on Research and Development in Information Retrieval}} (New Orleans, Louisiana, USA) \emph{(\bibinfo{series}{SIGIR '01})}. \bibinfo{publisher}{Association for Computing Machinery}, \bibinfo{address}{New York, NY, USA}, \bibinfo{pages}{120–127}.
\newblock
\showISBNx{1581133316}
\href{https://doi.org/10.1145/383952.383972}{doi:\nolinkurl{10.1145/383952.383972}}


\bibitem[Li et~al\mbox{.}(2024)]%
        {cross_qe-24}
\bibfield{author}{\bibinfo{person}{Minghan Li}, \bibinfo{person}{Honglei Zhuang}, \bibinfo{person}{Kai Hui}, \bibinfo{person}{Zhen Qin}, \bibinfo{person}{Jimmy Lin}, \bibinfo{person}{Rolf Jagerman}, \bibinfo{person}{Xuanhui Wang}, {and} \bibinfo{person}{Michael Bendersky}.} \bibinfo{year}{2024}\natexlab{}.
\newblock \showarticletitle{Can Query Expansion Improve Generalization of Strong Cross-Encoder Rankers?}. In \bibinfo{booktitle}{\emph{Proceedings of the 47th International ACM SIGIR Conference on Research and Development in Information Retrieval}} (Washington DC, USA) \emph{(\bibinfo{series}{SIGIR '24})}. \bibinfo{publisher}{Association for Computing Machinery}, \bibinfo{address}{New York, NY, USA}, \bibinfo{pages}{2321–2326}.
\newblock
\showISBNx{9798400704314}
\href{https://doi.org/10.1145/3626772.3657979}{doi:\nolinkurl{10.1145/3626772.3657979}}


\bibitem[Lin et~al\mbox{.}(2021)]%
        {pyserini-21}
\bibfield{author}{\bibinfo{person}{Jimmy Lin}, \bibinfo{person}{Xueguang Ma}, \bibinfo{person}{Sheng-Chieh Lin}, \bibinfo{person}{Jheng-Hong Yang}, \bibinfo{person}{Ronak Pradeep}, {and} \bibinfo{person}{Rodrigo Nogueira}.} \bibinfo{year}{2021}\natexlab{}.
\newblock \showarticletitle{Pyserini: A Python Toolkit for Reproducible Information Retrieval Research with Sparse and Dense Representations}. In \bibinfo{booktitle}{\emph{Proceedings of the 44th International ACM SIGIR Conference on Research and Development in Information Retrieval}} (Virtual Event, Canada) \emph{(\bibinfo{series}{SIGIR '21})}. \bibinfo{publisher}{Association for Computing Machinery}, \bibinfo{address}{New York, NY, USA}, \bibinfo{pages}{2356–2362}.
\newblock
\showISBNx{9781450380379}
\href{https://doi.org/10.1145/3404835.3463238}{doi:\nolinkurl{10.1145/3404835.3463238}}


\bibitem[Mackie et~al\mbox{.}(2023a)]%
        {mackie2023generativepseudorelevantfeedbacksparse}
\bibfield{author}{\bibinfo{person}{Iain Mackie}, \bibinfo{person}{Shubham Chatterjee}, {and} \bibinfo{person}{Jeffrey Dalton}.} \bibinfo{year}{2023}\natexlab{a}.
\newblock \bibinfo{title}{Generative and Pseudo-Relevant Feedback for Sparse, Dense and Learned Sparse Retrieval}.
\newblock
\showeprint[arxiv]{2305.07477}~[cs.IR]
\urldef\tempurl%
\url{https://arxiv.org/abs/2305.07477}
\showURL{%
\tempurl}


\bibitem[Mackie et~al\mbox{.}(2023b)]%
        {GRF-23}
\bibfield{author}{\bibinfo{person}{Iain Mackie}, \bibinfo{person}{Shubham Chatterjee}, {and} \bibinfo{person}{Jeffrey Dalton}.} \bibinfo{year}{2023}\natexlab{b}.
\newblock \showarticletitle{Generative Relevance Feedback with Large Language Models}. In \bibinfo{booktitle}{\emph{Proceedings of the 46th International ACM SIGIR Conference on Research and Development in Information Retrieval}} (Taipei, Taiwan) \emph{(\bibinfo{series}{SIGIR '23})}. \bibinfo{publisher}{Association for Computing Machinery}, \bibinfo{address}{New York, NY, USA}, \bibinfo{pages}{2026–2031}.
\newblock
\showISBNx{9781450394086}
\href{https://doi.org/10.1145/3539618.3591992}{doi:\nolinkurl{10.1145/3539618.3591992}}


\bibitem[Nguyen et~al\mbox{.}(2022)]%
        {legal-retrieval-22}
\bibfield{author}{\bibinfo{person}{Ha-Thanh Nguyen}, \bibinfo{person}{Manh-Kien Phi}, \bibinfo{person}{Xuan-Bach Ngo}, \bibinfo{person}{Vu Tran}, \bibinfo{person}{Le-Minh Nguyen}, {and} \bibinfo{person}{Minh-Phuong Tu}.} \bibinfo{year}{2022}\natexlab{}.
\newblock \showarticletitle{Attentive deep neural networks for legal document retrieval}.
\newblock \bibinfo{journal}{\emph{Artif. Intell. Law}} \bibinfo{volume}{32}, \bibinfo{number}{1} (\bibinfo{date}{Dec.} \bibinfo{year}{2022}), \bibinfo{pages}{57–86}.
\newblock
\showISSN{0924-8463}
\href{https://doi.org/10.1007/s10506-022-09341-8}{doi:\nolinkurl{10.1007/s10506-022-09341-8}}


\bibitem[Nigam et~al\mbox{.}(2019)]%
        {product-19}
\bibfield{author}{\bibinfo{person}{Priyanka Nigam}, \bibinfo{person}{Yiwei Song}, \bibinfo{person}{Vijai Mohan}, \bibinfo{person}{Vihan Lakshman}, \bibinfo{person}{Weitian~(Allen) Ding}, \bibinfo{person}{Ankit Shingavi}, \bibinfo{person}{Choon~Hui Teo}, \bibinfo{person}{Hao Gu}, {and} \bibinfo{person}{Bing Yin}.} \bibinfo{year}{2019}\natexlab{}.
\newblock \showarticletitle{Semantic Product Search}. In \bibinfo{booktitle}{\emph{Proceedings of the 25th ACM SIGKDD International Conference on Knowledge Discovery \& Data Mining}} (Anchorage, AK, USA) \emph{(\bibinfo{series}{KDD '19})}. \bibinfo{publisher}{Association for Computing Machinery}, \bibinfo{address}{New York, NY, USA}, \bibinfo{pages}{2876–2885}.
\newblock
\showISBNx{9781450362016}
\href{https://doi.org/10.1145/3292500.3330759}{doi:\nolinkurl{10.1145/3292500.3330759}}


\bibitem[OpenAI(2024)]%
        {openai2024gpt41}
\bibfield{author}{\bibinfo{person}{OpenAI}.} \bibinfo{year}{2024}\natexlab{}.
\newblock \bibinfo{title}{GPT-4.1}.
\newblock \bibinfo{howpublished}{\url{https://openai.com/index/gpt-4-1/}}.
\newblock


\bibitem[Roberts et~al\mbox{.}(2021)]%
        {trec-covid-21}
\bibfield{author}{\bibinfo{person}{Kirk Roberts}, \bibinfo{person}{Tasmeer Alam}, \bibinfo{person}{Steven Bedrick}, \bibinfo{person}{Dina Demner-Fushman}, \bibinfo{person}{Kyle Lo}, \bibinfo{person}{Ian Soboroff}, \bibinfo{person}{Ellen Voorhees}, \bibinfo{person}{Lucy~Lu Wang}, {and} \bibinfo{person}{William~R. Hersh}.} \bibinfo{year}{2021}\natexlab{}.
\newblock \showarticletitle{Searching for scientific evidence in a pandemic: An overview of TREC-COVID}.
\newblock \bibinfo{journal}{\emph{J. of Biomedical Informatics}} \bibinfo{volume}{121}, \bibinfo{number}{C} (\bibinfo{date}{Sept.} \bibinfo{year}{2021}), \bibinfo{numpages}{16}~pages.
\newblock
\showISSN{1532-0464}
\href{https://doi.org/10.1016/j.jbi.2021.103865}{doi:\nolinkurl{10.1016/j.jbi.2021.103865}}


\bibitem[Robertson(1991)]%
        {qe-91}
\bibfield{author}{\bibinfo{person}{S.~E. Robertson}.} \bibinfo{year}{1991}\natexlab{}.
\newblock \showarticletitle{On term selection for query expansion}.
\newblock \bibinfo{journal}{\emph{J. Doc.}} \bibinfo{volume}{46}, \bibinfo{number}{4} (\bibinfo{date}{Jan.} \bibinfo{year}{1991}), \bibinfo{pages}{359–364}.
\newblock
\showISSN{0022-0418}
\href{https://doi.org/10.1108/eb026866}{doi:\nolinkurl{10.1108/eb026866}}


\bibitem[Robertson et~al\mbox{.}(1994)]%
        {bm25}
\bibfield{author}{\bibinfo{person}{Stephen~E. Robertson}, \bibinfo{person}{Steve Walker}, \bibinfo{person}{Susan Jones}, \bibinfo{person}{Micheline Hancock-Beaulieu}, {and} \bibinfo{person}{Mike Gatford}.} \bibinfo{year}{1994}\natexlab{}.
\newblock \showarticletitle{Okapi at TREC-3}. In \bibinfo{booktitle}{\emph{Text Retrieval Conference}}.
\newblock
\urldef\tempurl%
\url{https://api.semanticscholar.org/CorpusID:41563977}
\showURL{%
\tempurl}


\bibitem[Shen et~al\mbox{.}(2024)]%
        {lamer-24}
\bibfield{author}{\bibinfo{person}{Tao Shen}, \bibinfo{person}{Guodong Long}, \bibinfo{person}{Xiubo Geng}, \bibinfo{person}{Chongyang Tao}, \bibinfo{person}{Yibin Lei}, \bibinfo{person}{Tianyi Zhou}, \bibinfo{person}{Michael Blumenstein}, {and} \bibinfo{person}{Daxin Jiang}.} \bibinfo{year}{2024}\natexlab{}.
\newblock \showarticletitle{Retrieval-Augmented Retrieval: Large Language Models are Strong Zero-Shot Retriever}. In \bibinfo{booktitle}{\emph{Findings of the Association for Computational Linguistics: ACL 2024}}, \bibfield{editor}{\bibinfo{person}{Lun-Wei Ku}, \bibinfo{person}{Andre Martins}, {and} \bibinfo{person}{Vivek Srikumar}} (Eds.). \bibinfo{publisher}{Association for Computational Linguistics}, \bibinfo{address}{Bangkok, Thailand}, \bibinfo{pages}{15933--15946}.
\newblock
\href{https://doi.org/10.18653/v1/2024.findings-acl.943}{doi:\nolinkurl{10.18653/v1/2024.findings-acl.943}}


\bibitem[Thakur et~al\mbox{.}(2021)]%
        {thakur2021beir}
\bibfield{author}{\bibinfo{person}{Nandan Thakur}, \bibinfo{person}{Nils Reimers}, \bibinfo{person}{Andreas R{\"u}ckl{\'e}}, \bibinfo{person}{Abhishek Srivastava}, {and} \bibinfo{person}{Iryna Gurevych}.} \bibinfo{year}{2021}\natexlab{}.
\newblock \showarticletitle{{BEIR}: A Heterogeneous Benchmark for Zero-shot Evaluation of Information Retrieval Models}. In \bibinfo{booktitle}{\emph{Thirty-fifth Conference on Neural Information Processing Systems Datasets and Benchmarks Track (Round 2)}}.
\newblock
\urldef\tempurl%
\url{https://openreview.net/forum?id=wCu6T5xFjeJ}
\showURL{%
\tempurl}


\bibitem[Touvron et~al\mbox{.}(2023)]%
        {llama-23}
\bibfield{author}{\bibinfo{person}{Hugo Touvron}, \bibinfo{person}{Thibaut Lavril}, \bibinfo{person}{Gautier Izacard}, \bibinfo{person}{Xavier Martinet}, \bibinfo{person}{Marie-Anne Lachaux}, \bibinfo{person}{Timothée Lacroix}, \bibinfo{person}{Baptiste Rozière}, \bibinfo{person}{Naman Goyal}, \bibinfo{person}{Eric Hambro}, \bibinfo{person}{Faisal Azhar}, \bibinfo{person}{Aurelien Rodriguez}, \bibinfo{person}{Armand Joulin}, \bibinfo{person}{Edouard Grave}, {and} \bibinfo{person}{Guillaume Lample}.} \bibinfo{year}{2023}\natexlab{}.
\newblock \bibinfo{title}{LLaMA: Open and Efficient Foundation Language Models}.
\newblock
\showeprint[arxiv]{2302.13971}~[cs.CL]
\urldef\tempurl%
\url{https://arxiv.org/abs/2302.13971}
\showURL{%
\tempurl}


\bibitem[Wang et~al\mbox{.}(2023)]%
        {query2doc-23}
\bibfield{author}{\bibinfo{person}{Liang Wang}, \bibinfo{person}{Nan Yang}, {and} \bibinfo{person}{Furu Wei}.} \bibinfo{year}{2023}\natexlab{}.
\newblock \showarticletitle{Query2doc: Query Expansion with Large Language Models}. In \bibinfo{booktitle}{\emph{Proceedings of the 2023 Conference on Empirical Methods in Natural Language Processing}}, \bibfield{editor}{\bibinfo{person}{Houda Bouamor}, \bibinfo{person}{Juan Pino}, {and} \bibinfo{person}{Kalika Bali}} (Eds.). \bibinfo{publisher}{Association for Computational Linguistics}, \bibinfo{address}{Singapore}, \bibinfo{pages}{9414--9423}.
\newblock
\href{https://doi.org/10.18653/v1/2023.emnlp-main.585}{doi:\nolinkurl{10.18653/v1/2023.emnlp-main.585}}


\bibitem[Weller et~al\mbox{.}(2024)]%
        {qe-fail-24}
\bibfield{author}{\bibinfo{person}{Orion Weller}, \bibinfo{person}{Kyle Lo}, \bibinfo{person}{David Wadden}, \bibinfo{person}{Dawn Lawrie}, \bibinfo{person}{Benjamin Van~Durme}, \bibinfo{person}{Arman Cohan}, {and} \bibinfo{person}{Luca Soldaini}.} \bibinfo{year}{2024}\natexlab{}.
\newblock \showarticletitle{When do Generative Query and Document Expansions Fail? A Comprehensive Study Across Methods, Retrievers, and Datasets}. In \bibinfo{booktitle}{\emph{Findings of the Association for Computational Linguistics: EACL 2024}}, \bibfield{editor}{\bibinfo{person}{Yvette Graham} {and} \bibinfo{person}{Matthew Purver}} (Eds.). \bibinfo{publisher}{Association for Computational Linguistics}, \bibinfo{address}{St. Julian{'}s, Malta}, \bibinfo{pages}{1987--2003}.
\newblock
\urldef\tempurl%
\url{https://aclanthology.org/2024.findings-eacl.134/}
\showURL{%
\tempurl}


\bibitem[Xiao et~al\mbox{.}(2024)]%
        {bge_embedding}
\bibfield{author}{\bibinfo{person}{Shitao Xiao}, \bibinfo{person}{Zheng Liu}, \bibinfo{person}{Peitian Zhang}, \bibinfo{person}{Niklas Muennighoff}, \bibinfo{person}{Defu Lian}, {and} \bibinfo{person}{Jian-Yun Nie}.} \bibinfo{year}{2024}\natexlab{}.
\newblock \showarticletitle{C-Pack: Packed Resources For General Chinese Embeddings}. In \bibinfo{booktitle}{\emph{Proceedings of the 47th International ACM SIGIR Conference on Research and Development in Information Retrieval}} (Washington DC, USA) \emph{(\bibinfo{series}{SIGIR '24})}. \bibinfo{publisher}{Association for Computing Machinery}, \bibinfo{address}{New York, NY, USA}, \bibinfo{pages}{641–649}.
\newblock
\showISBNx{9798400704314}
\href{https://doi.org/10.1145/3626772.3657878}{doi:\nolinkurl{10.1145/3626772.3657878}}


\bibitem[Xiong et~al\mbox{.}(2021)]%
        {ance-20}
\bibfield{author}{\bibinfo{person}{Lee Xiong}, \bibinfo{person}{Chenyan Xiong}, \bibinfo{person}{Ye Li}, \bibinfo{person}{Kwok-Fung Tang}, \bibinfo{person}{Jialin Liu}, \bibinfo{person}{Paul~N. Bennett}, \bibinfo{person}{Junaid Ahmed}, {and} \bibinfo{person}{Arnold Overwijk}.} \bibinfo{year}{2021}\natexlab{}.
\newblock \showarticletitle{Approximate Nearest Neighbor Negative Contrastive Learning for Dense Text Retrieval}. In \bibinfo{booktitle}{\emph{International Conference on Learning Representations}}.
\newblock
\urldef\tempurl%
\url{https://openreview.net/forum?id=zeFrfgyZln}
\showURL{%
\tempurl}


\bibitem[Zhang et~al\mbox{.}(2024)]%
        {mugi-24}
\bibfield{author}{\bibinfo{person}{Le Zhang}, \bibinfo{person}{Yihong Wu}, \bibinfo{person}{Qian Yang}, {and} \bibinfo{person}{Jian-Yun Nie}.} \bibinfo{year}{2024}\natexlab{}.
\newblock \showarticletitle{Exploring the Best Practices of Query Expansion with Large Language Models}. In \bibinfo{booktitle}{\emph{Findings of the Association for Computational Linguistics: EMNLP 2024}}, \bibfield{editor}{\bibinfo{person}{Yaser Al-Onaizan}, \bibinfo{person}{Mohit Bansal}, {and} \bibinfo{person}{Yun-Nung Chen}} (Eds.). \bibinfo{publisher}{Association for Computational Linguistics}, \bibinfo{address}{Miami, Florida, USA}, \bibinfo{pages}{1872--1883}.
\newblock
\href{https://doi.org/10.18653/v1/2024.findings-emnlp.103}{doi:\nolinkurl{10.18653/v1/2024.findings-emnlp.103}}


\end{thebibliography}

\end{document}